\documentclass[twocolumn,aps,prb]{revtex4}

\usepackage[dvips]{graphicx}
\usepackage{epsfig}
\usepackage{amssymb}

\newcommand{\ftdB}{f_\mathrm{3dB}}
\newcommand{\Vb}{V_\mathrm{b}}
\newcommand{\VbMQW}{V_\mathrm{b,MQW}}
\newcommand{\Rlat}{R_\mathrm{lat}}
\newcommand{\RMQW}{R_\mathrm{MQW}}
\newcommand{\Vpp}{V_\mathrm{pp}}

\begin{document}

\title{Realization of efficient electroabsorption using intersubband transitions in step quantum wells}

\author{P. Holmstr\"om,$^1$\footnote{Electronic mail: petterh@kth.se} P. J\"anes,$^{1,2}$ U. Ekenberg$^1$
and L. Thyl\'en$^1$}
\affiliation{$^1$Laboratory of Photonics and Microwave Engineering, Department of Microelectronics and
Applied Physics,\\Royal Institute of Technology (KTH), SE-164 40 Kista, Sweden.\\
$^2$Present address: Proximion Fiber Systems AB, SE-16440 Kista,
Sweden.}

\date{\today}

\begin{abstract}
We have demonstrated efficient intersubband electroabsorption in
InGaAs/InAlGaAs/InAlAs step quantum wells grown by metal-organic
vapor phase epitaxy (MOVPE). An absorption modulation of 6 dB at
$\lambda=6.0\ \mu$m due to Stark shift of the IS absorption was
achieved at a low voltage swing of $\pm 0.5$ V in a multipass
waveguide structure. Based on the experimental results it is
estimated that an electroabsorption modulator with a low
peak-to-peak voltage of $\Vpp=0.9$ V can yield a modulation speed of
$\ftdB\approx 120$ GHz with the present material by using a strongly
confining surface plasmon waveguide of 30 $\mu$m length.
\end{abstract}

\maketitle



Intersubband (IS) transitions have primarily been applied to
quantum-well infrared photodetectors (QWIPs) \cite{liu:00} and
quantum cascade lasers (QCLs).\cite{faist:02} However, the
properties of IS transitions are also attractive for application to
electroabsorption (EA) modulators. A fundamental advantage of
intersubband transitions is that the electron subbands are
essentially parallel to each other. The weak dependence on the
in-plane wave vector results in sharply peaked IS resonances and a
potential for strong absorption implying compact modulators. The
broadening of the peaks due to subband nonparabolicity has been
shown to be effectively cancelled by collective
phenomena.\cite{warburton:00} A small IS absorption linewidth
$\Gamma$ is imperative for a high RC-limited speed in IS-based
EA-modulators, as generally the modulator capacitance
$C\sim\Gamma^3$.[\onlinecite{holmstrom:01b}] Other advantages are
the rapid IS relaxation time of only $\sim 1$ ps which enables
saturation resistant modulators, and the possibility to achieve a
negative chirp parameter.

IS based modulation has previously been demonstrated at mid-IR
wavelengths,\cite{karunasiri:90,liu:03} however using rather large
voltages. In this letter we show experimentally that an absorption
modulation of 6 dB at $\lambda=6.0\ \mu$m due to Stark shift of the
IS absorption can be achieved with a voltage swing as low as $\pm
0.5$ V. Based on these results we estimate that an electroabsorption
modulator with a low peak-to-peak voltage of $\Vpp=0.9$ V can yield
a modulation speed of $\ftdB\approx 120$ GHz with the present
material by using a strongly confining surface plasmon waveguide of
30 $\mu$m length.

\begin{figure}[!t]
  \begin{center}
     \epsfig{file=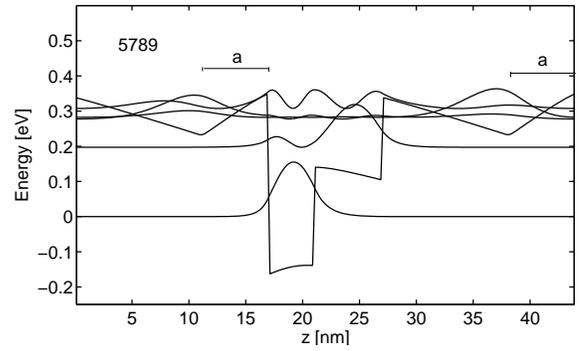,width=76mm}
  \end{center}
  \vspace{-3ex}
  \caption{Potential profile of the step QW in the sample 5789 with no applied bias voltage. Also shown are the
  moduli squared of the bound states in the step QW as well as two mostly barrier confined states. The
  samples differ by the thickness $a$ of the spacer layer between the $\delta$-doping layer and the InGaAs QW
  layer. $a=5.6$ nm and 8.3 nm in the samples 5789 and 5791, respectively.}
  \label{fig:QW_5789}
\end{figure}

The samples were grown in an Aixtron AIX-200 MOVPE system on
semi-insulating InP:Fe substrates. First a 500 nm n-InP
($5\times10^{17}$ cm$^{-3}$) buffer layer was grown. It was followed
by a 60 nm lattice-matched n-AlGaInAs ($x_{Al}$=0.37,
$5\times10^{17}$ cm$^{-3}$) layer, which was grown immediately
before the MQW layer, with a similar one just after the MQW. The
purpose of these two n-AlGaInAs layers is to line up the Fermi
energies inside and outside the MQW without bend
bending,\cite{holmstrom:01b} in order to obtain the same IS
transition energies in all step QWs. On top of the upper 60 nm
n-AlGaInAs layer a 550 nm n-InP ($5\times10^{17}$ cm$^{-3}$) cap
layer was grown. Each step QW consists of a 4.0 nm InGaAs well
layer, a 6.0 nm AlGaInAs step layer, separated by 16.7 nm InAlAs
barrier layers. The barriers were Si $\delta$-doped at
$n_\mathrm{D}=2.2\times10^{12}$ cm$^{-2}$ to supply electrons in the
step QWs. The $\delta$-dopings were performed by halting the growth
while supplying SiH$_4$, whereafter the growth proceeded without
further delay. Each sample had an MQW comprising 10 step QWs. Two
samples 5789 and 5791 were grown which differed by the position of
the $\delta$-doping layer. The potential profile of one step QW in
5789 without any applied bias voltage is shown in Fig.
\ref{fig:QW_5789}.

Structural characterization of the grown samples was performed by a
combination of high-resolution X-ray diffraction (XRD) measurements
and simulation of the two observed IS resonance peak energies
$\tilde{E}_{12}$ and $\tilde{E}_{13}$. By XRD $2\theta-\omega$ scans
the MQW period $L_p=26.7\pm0.2$ nm in the samples could be reliably
obtained. But since the MQW comprises a repetition of three
different and nearly lattice-matched layers, viz. the InGaAs well
layer, the InAlGaAs step layer and the InAlAs barrier, the
individual thicknesses and compositions of these were not available
from XRD results alone. However the IS peak energies
$\tilde{E}_{12}$ and $\tilde{E}_{13}$ as well as the ratio between
the two oscillator strengths $f_{13}/f_{12}$ adds enough information
to determine the thickness of the well and step layers as well as
the conduction band edge of the step layer corresponding to a
composition of Al$_{0.30}$Ga$_{0.18}$In$_{0.52}$As.

The IS transition energies were modeled in the envelope function
approximation. The structures examined here in many respects
coincide with the one we previously evaluated in a modulator
simulation.\cite{holmstrom:01b} We have considered the
nonparabolicity of the conduction band.\cite{sirtori:94} We solved
for the step QW potential self-consistently including the Hartree
space-charge potential and an exchange-correlation
potential.\cite{hedin:71,bloss:89} The exchange-correlation effects
were thus included as a one-particle exchange-correlation potential.
Further we also considered the collective shift of the IS resonance
peak due to the depolarization and exciton effects.\cite{ando:77}
Material parameters were obtained from the review paper by
Vurgaftman \emph{et al}.\cite{vurgaftman:01}

The IS absorption was characterized in a multi-pass geometry in the
usual fashion where two bevelled edges and the back side of a
$5\times7$ mm piece were polished to allow multiple internal
reflections. The spectra were measured using Fourier-transform
infrared (FTIR) spectroscopy, and isolation of the IS absorbance was
achieved by taking the ratio of the transmittance spectra of TM and
TE-polarized light, $A=-\log_{10}(T_{\mathrm TM}/T_{\mathrm TE})$.
In addition, a smooth background offset due to the polarization
dependence in the FTIR spectrometer was subtracted. Due to
polarization selection rules IS transitions almost exclusively
couple to TM-polarized light. No significant absorption was observed
for TE-polarized light.

\begin{figure}[!tb]
  \begin{center}
     \epsfig{file=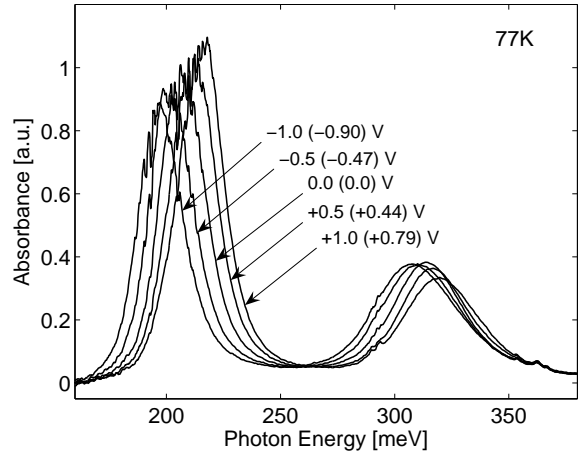,width=86mm}
  \end{center}
  \vspace{-3ex}
  \caption{Intersubband absorption spectra measured in sample 5789 for different applied voltages at T=77 K.
  Clear Stark shifts of the $1 \rightarrow 2$ and $1 \rightarrow 3$ IS resonances are observed. Indicated for
  each spectrum is the applied voltage $\Vb$ and within parentheses the voltage over the MQW $\VbMQW$.}
  \label{fig:Stark_shift_spectra}
\end{figure}

To enable voltage to be applied over the MQWs Ti/Pt/Au contacts were
deposited on the n-InP cap layer and on the buffer n-InP layer below
the MQW stack. The bottom contact was deposited after lithography
and dry-etching down to the bottom contact layer. The contacts were
performed as stripes $5\times0.2$ mm$^2$ with a distance of 4 mm
between them. At zero bias the peak widths did not differ much
between 77 K and room temperature. Clear Stark shifts of the $1
\rightarrow 2$ and $1 \rightarrow 3$ IS resonances were observed at
low applied voltages, Fig. \ref{fig:Stark_shift_spectra}. The EA was
characterized while keeping the samples in a liquid nitrogen
cryostat at T=77 K. Although the envisaged $\mu$m-scale ridge
waveguide device should be operated at room-temperature, EA
characterization at low temperature is necessary due to the large
area of the multipass device. IV-characterization still revealed a
significant leakage current. To obtain the part of the applied bias
voltage $\Vb$ that was applied over the MQW it is thus necessary to
account for the voltage drop in the contact layers. We model this by
considering that the device resistance $R$ is composed of two parts
in series $R=\Rlat+\RMQW$. The resistance $\Rlat$ is associated with
the lateral conduction in the n-InP contact layers and laterally in
the MQW. We assume that $\Rlat$ is independent of the applied bias.
The voltage that is applied over the MQW is then obtained by voltage
division, i.e. $\VbMQW=\Vb(R-\Rlat)/R$. The bias dependent multipass
device resistance $R=\Vb/I$, where $I$ is the current through the
device.

\begin{figure}[!t]
  \begin{center}
     \epsfig{file=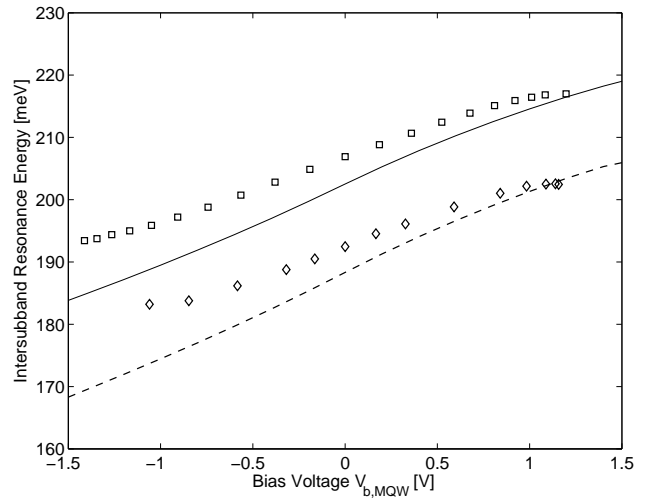,width=84mm}
  \end{center}
  \vspace{-3ex}
  \caption{Stark shift of the $1 \rightarrow 2$ intersubband absorption peak position vs. the voltage
  $\VbMQW$ applied over the MQW. The symbols indicate experimental IS peak positions in the samples 5789
  (squares) and 5791 (diamonds). The solid and dashed lines are the
  calculated Stark shift in the samples 5789 and 5791,
  respectively.}
  \label{fig:Stark_shift_calc_vs_meas}
\end{figure}

The shift of the $1 \rightarrow 2$ IS resonance due to the Stark
effect in both the samples is shown in Fig.
\ref{fig:Stark_shift_calc_vs_meas}. The voltage drop that occurred
in the contact layers was compensated for by plotting the IS
absorption peak positions vs. the voltage over the MQW, $\VbMQW$.
The experimental Stark shifts were still about 20\% smaller than the
calculated ones for sharp interfaces. Possible causes of the
difference between the experimental and calculated Stark shifts are
the uncertainty about the voltage drop outside the quantum well and
intermixing of the QW interfaces. IS based structures have generally
been grown by molecular-beam epitaxy (MBE) owing to its better
control of small layer thicknesses and lower growth temperatures
which should reduce interface intermixing. The present structures
were, however, grown by metal-organic vapor-phase epitaxy (MOVPE)
which is considered more suitable for commercial production.

Following ref.[\onlinecite{holmstrom:01b}] we can predict the
performance of high-speed modulators. Instead of a multi-pass
geometry we then consider a surface plasmon waveguide. By having a
heavily doped semiconductor layer below the MQW structure and Au
above it we predicted that a modulator can be made as short as 30
$\mu$m. In simulations for sharp interfaces we have predicted very
high $RC$-limited room-temperature modulation speeds $\ftdB$ at low
voltages,\cite{holmstrom:01b,janes:02} e.g. $\ftdB\sim190$ GHz at a
driving voltage of only $\Vpp=0.9$ V \onlinecite{holmstrom:01b}.
Based on the experimental results in this paper it should still be
possible to reach $\ftdB\sim120$ GHz.

High-speed IS modulators operating at $\lambda=6.0\ \mu$m can be
useful for interconnects with Si waveguides. Si has a transmission
window from 1.2 to 6.5 $\mu$m.[\onlinecite{jalali:06}] The
atmospheric absorption at the present wavelength is rather large.
But with a slightly different design it should be no problem to
reach the windows at $\lambda=3 - 5\ \mu$m and $\lambda= 8-12\
\mu$m. Such modulators would be very valuable for free-space
communication.\cite{killinger:02} This is an area of increasing
importance. In addition to low absorption at these wavelengths in
clean air, the absorption due to water droplets, oil droplets and
dust have recently been studied in wind tunnels.\cite{martini:04}
The absorption was found to be an order of magnitude or more lower
than for the commonly used wavelengths of 850, 1300 and 1550 nm. The
present predicted speed is much higher than for directly modulated
quantum cascade lasers. Continuous wave operation of a quantum
cascade laser at room temperature has been demonstrated at $\lambda
= 9.2 \, \mu$m.[\onlinecite{beck:02}] When it comes to interband
modulators at mid-IR wavelengths it appears that the research
activities have been quite low so far.

The concept is also very promising for fiber-optical communication
wavelengths like $\lambda=1.55\ \mu$m. This implementation requires
material pairs with very large conduction band offsets. Such
materials are generally less mature resulting in clearly larger line
widths $\Gamma$, but with continued improvement of sample quality
they should become superior to present interband
modulators.\cite{holmstrom:06}

In conclusion we have measured efficient electroabsorption in
InP-based step quantum well structures. Clear Stark shifts were
observed with a bias of the order 1 V. The results indicate that a
30 $\mu$m long modulator with a surface plasmon waveguide can reach
a speed exceeding $\ftdB\sim100$ GHz.

The authors are grateful to A. Patel for growing the samples by
MOVPE. We thank St\'{e}phane Junique, Lars Sj\"{o}kvist and Jens
Zander for valuable discussions.


\end{document}